# Precise Determination of Minimum Achievable Temperature for Solid-State Optical Refrigeration

**Denis V. Seletskiy[a,c,*], Seth D. Melgaard[a], Richard I. Epstein[a], Alberto Di Lieto[b], Mauro Tonelli[b], Mansoor Sheik-Bahae[a]**

[a] *University of New Mexico, Physics and Astronomy Department, Albuquerque NM 87131, USA*

[b] *NEST Istituto Nanoscienze- CNR, Dipartimento di Fisica Università di Pisa, Largo B. Pontecorvo 3, 56127 Pisa, Italy.*

[c] *Air Force Research Laboratory, Space Vehicles Directorate, Kirtland AFB, NM 87117, USA*

**Abstract:** We measure the minimum achievable temperature (MAT) as a function of excitation wavelength in anti-Stokes fluorescence cooling of high purity $Yb^{3+}$-doped $LiYF_4$ (Yb:YLF) crystal. Such measurements were obtained by developing a sensitive noncontact thermometry that is based on a two-band differential luminescence spectroscopy using balanced photo-detectors. These measurements are in excellent agreement with the prediction of the laser cooling model and identify MAT of 110 K at 1020 nm, corresponding to E4-E5 Stark manifold transition in Yb:YLF crystal.

[*] Corresponding author: d.seletskiy@gmail.com.

## 1. Introduction

The physical principle of optical refrigeration is based on a phonon-assisted anti-Stokes fluorescence. Spectrally narrow-band low-energy excitation photons produce energetically upshifted incoherent fluorescence emission, extracting heat away from the lattice in the process and resulting in cooling of the latter [1]. Optical refrigeration was first postulated by Pringsheim in 1929 [2], put on a solid thermodynamic footing by Landau in 1946 [3] and finally demonstrated in solids by Epstein and co-workers in 1995 [4]. First observation consisted of bulk cooling of an ytterbium-doped fluorozirconate glass (Yb:ZBLAN) by 0.3 K, starting from room temperature. Cooling of this material system has progressed over the years and culminated with demonstration of cooling to 208 K by 2005 [5]. In parallel with these results, other trivalent ions of Tm and Er were cooled on various transitions and in a wide variety of hosts (see Refs. [6,7] for recent reviews of this field). Interesting applications of optical refrigeration have also been proposed, including solid-state cryogenic refrigerators [8,9,1] and radiationally-balanced lasers [10]. Advancing toward realization of the former, laser cooling to 155 K (from room temperature) was recently demonstrated [11], followed by with cooling a semiconductor payload to 165 K [12], utilizing 5 mol % ytterbium doped lithium-lithium fluoride crystal (Yb:YLF). These bulk cooling results provided the proof-of-principle demonstration of operation of all-solid-state optical refrigerator at temperatures below what conventional Peltier devices can achieve.

In this paper we discuss rate-equation based cooling efficiency model [13,6] along with predictions of minimum achievable temperature (MAT) of 110 K at 1020 nm [11], corresponding to E4-E5 Stark manifold transition in Yb:YLF. In order to make these predictions, model is supplemented with experimentally determined quantities, details of which are discussed below. To verify these model predictions, we developed a highly sensitive differential spectroscopic technique that allows us to fully characterize cooling sample performance. Below we discuss several implementations and details of this technique along with recent results on Yb:YLF verifying MAT of 110 ± 5 K at 1020 nm, in excellent agreement with the model.

## 2. Model

The cooling efficiency is defined as the ratio of the cooling power ($P_{cool}$) to the absorbed laser power ($P_{abs}$), is given by [13]:

$$\eta_c(\lambda, T) = \frac{P_{cool}}{P_{abs}} = p(\lambda, T)\frac{\lambda}{\lambda_f(T)} - 1,$$ (1)

where $\lambda_f(T)$ is a temperature-dependent external mean fluorescence wavelength (i.e. including fluorescence trapping and reabsorption). The term $p(\lambda, T)$ is a probability of the conversion of a low-energy excitation photon into an escaped fluorescence photon:

$$p(\lambda, T) = \eta_{ext}\frac{\alpha(\lambda, T)}{\alpha(\lambda, T) + \alpha_b},$$ (2)

where $\eta_{ext}$ is the external quantum efficiency (EQE) defined as the fraction of excited ions that lead to a fluorescence photon exiting the host material; $\alpha(\lambda, T)$ is the resonant absorption of the ions while parasitic absorption on impurities is represented by the background absorption term $\alpha_b$. In the sign convention adopted here, positive $\eta_c$ corresponds to cooling. High $\eta_{ext}$ (> 99%) is typical for metastable f-f transition in rare-earth ions doped in hosts with low phonon energy (such as fluorides). In addition, $\eta_{ext}$ is largely temperature independent [14]. The remaining term on the right hand side of Eq. (2) represents the fraction of the absorbed photons by the rare-earth ions (e.g. $^2F_{7/2} - ^2F_{5/2}$ transition in Yb$^{3+}$), and is termed absorption efficiency $\eta_{abs}$. The origin of background absorption $\alpha_b$ is due to unwanted growth contamination (estimated to be much less than part per million), and is typically attributed to the presence of transition metals such as copper and iron, similar to glass hosts [15]. Following this assumption, $\alpha_b$ is also taken to be temperature independent, and broadband (i.e. independent of wavelength) within the spectral region of the cooling transition [15]. Thus, to obtain a quantitative analysis of the cooling efficiency, we perform independent experiments to evaluate the material-dependent constituents of Eq. (1), namely: $\eta_{ext}$, $\alpha_b$, $\alpha(\lambda, T)$ and $\lambda_f(T)$.

With decrease of temperature, positive (in the cooling region) cooling efficiency decreases due to diminishing resonant absorption $\alpha$ and red-shifting of $\lambda_f$, until it switches sign, corresponding to overall heating. The temperature corresponding to this cooling-to-heating transition is the minimum achievable temperature (MAT) for a given excitation wavelength (i.e. $\alpha$). A global MAT is defined as the minimum temperature $T_{min}$ for which $\eta_c(\lambda, T_{min}) = 0$, or $min(MAT(\lambda))$. Spectrum of MAT($\lambda$) thus uniquely characterizes laser cooling performance of a given material and hence is of great interest to access experimentally. The following sections detail the experimental procedures that allow us to accurately measure MAT spectrum as well as the four input parameters needed to evaluate the cooling efficiency of Eq. (1).

## 3. Experimental results and discussion

Before we discuss direct measurements of MAT($\lambda$), we point out that predictive capabilities of the cooling efficiency model (Eq. (1)) can be gained only if its individual components are well characterized, namely: $\eta_{ext}$, $\alpha_b$, $\alpha(\lambda, T)$ and $\lambda_f(T)$.

### 3.1 Spectroscopic data for Yb:YLF

We conduct our studies on high purity Czochralski-grown 5 mol % doped Yb:YLF crystal [16] of dimensions $3 \times 3 \times 9$ mm³, used previously for bulk cryocooling results [11]. Polarized and unpolarized instrument-response-corrected fluorescence spectra of the crystal are obtained versus temperature from 100 K to 300 K (in 10 K increments), inside of a liquid-Nitrogen cryostat. Figure (1a) shows temperature dependent fluorescence spectra, as normalized by integrated value at 100 K. Reciprocity analysis [17] is then performed to calculate $\alpha(\lambda, T)$ (Fig. (1b)). Effects of reabsorption become significant, especially at low temperatures. For modeling of the cooling efficiency, we reproduce the exact geometry of the cooling experiments, namely exciting the beam through the center axis along the largest dimension $L = 9$ mm of the crystal. Mean fluorescence wavelength $\lambda_f(T)$ is calculated by taking a first moment of the unpolarized fluorescence spectra (Fig. (1c)), exhibiting approximately linear dependence where a fit to $\lambda_f(T) = mT + b$, with $m = -0.031 \pm 0.001$ nm/K and $b = 1008.9 \pm 0.1$ nm was obtained.

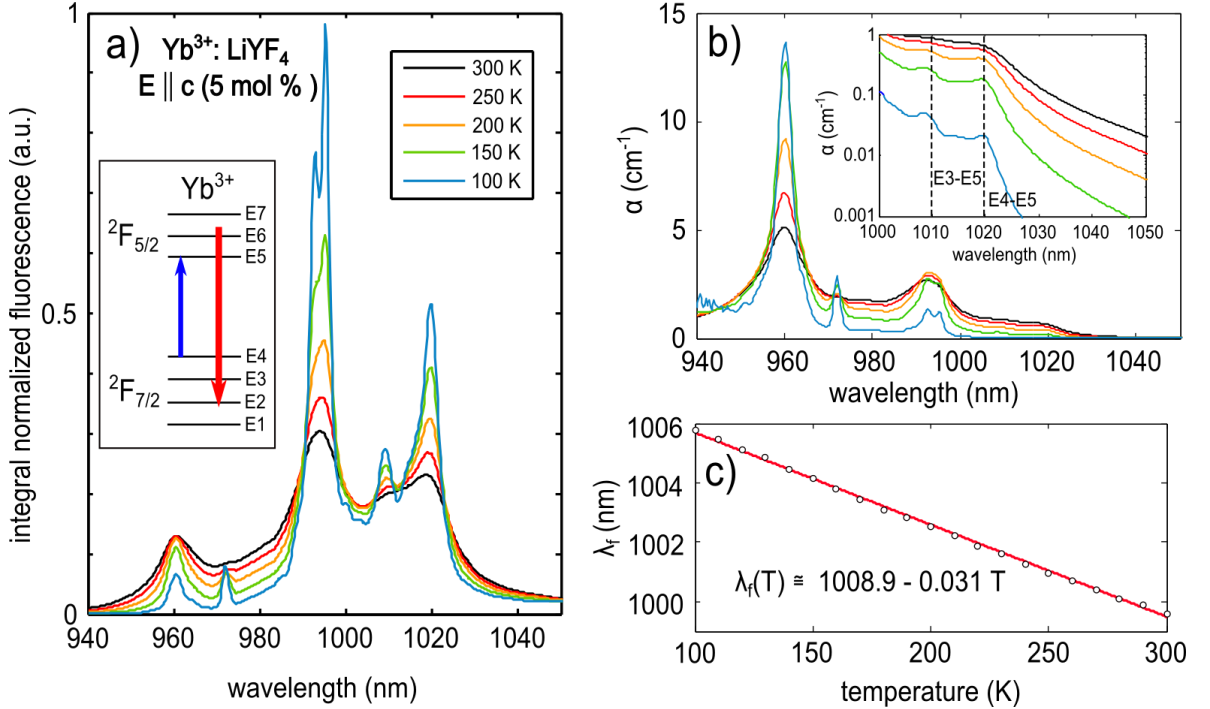

**Figure 1:** (a) Temperature-dependent fluorescence spectra of Yb³⁺:LiYF₄ crystal in E∥c orientation (inset shows Stark manifolds of Yb³⁺ $^2F_{5/2} - {}^2F_{7/2}$ transition) normalized by integrated value at 100 K; (b) temperature-dependent absorption spectra of a Yb:YLF (5 mol %) crystal with same color coding as panel a and also for E∥c orientation; inset shows anti-Stokes absorption (cooling) tail on a semi-logarithmic scale, with resonant features corresponding to E3-E5 and E4-E5 Stark manifold transitions; (c) temperature dependence of the mean fluorescence wavelength $\lambda_f(T)$ along with an approximate linear fit in the temperature range of $100 - 300$ K (see text for more details).

As was mentioned in previous section, external quantum efficiency $\eta_{ext}$ is typically independent of temperature [14]. Similarly, the background absorption coefficient $\alpha_b$ is also assumed to be nearly independent of temperature. Therefore, these two quantities are only determined in experiments conducted at room temperature. The validity of these assumptions however will be tested by the outcome of the experiments reported below.

A high purity Czochralski-grown 5% doped Yb:YLF crystals of dimensions $3\times3\times9$ mm$^3$ is situated in ambient air on thin microscope coverslips (to minimize conductive load) and is excited in E∥c orientation via a tunable CW Ti:Sapphire laser (950nm-1080nm, 1.3-1.8 W). Subsequent temperature change is monitored using a calibrated bolometric thermal camera. Recall that $\eta_c = P_{cool}/P_{abs}$ with $P_{cool} \cong C\Delta T$ and $P_{abs} = P_0[1 - exp(-\alpha(\lambda)L)]$, where $C$ is thermal-load dependent constant, $\Delta T$ is the laser-induced temperature change and $P_0$ is incident power on the crystal. The measured quantity $\Delta T/P_{abs}$ is thus proportional to the cooling efficiency. Figure 2 plots scaled measured values (of the cooling efficiency with black circles) versus the excitation wavelength. A fit, based on Eq. (1), yields values of $\eta_{ext} = 0.995 \pm 0.001$ and $\alpha_b = (4.0 \pm 0.2)\times10^{-4}$ cm$^{-1}$.

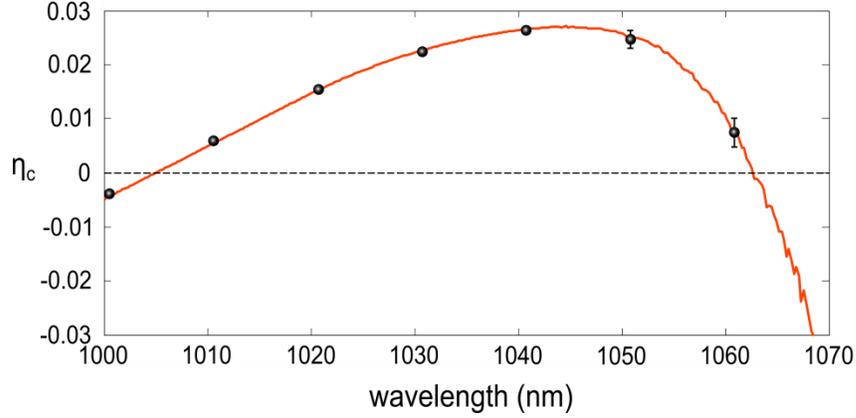

**Figure 2:** Measurement of the cooling efficiency at 300 K. The fit (red line) of the data (circles) is fit based on Eq. (1) and yields $\eta_{ext} = 0.995 \pm 0.001$ and $\alpha_b = (4.0 \pm 0.2)\times10^{-4}$ cm$^{-1}$, with $\alpha(\lambda,T=300K)$ and $\lambda_f(T=300K)$, determined as described in the text.

We also note, that similar analysis of $\eta_{ext}$, $\alpha_b$, $\alpha(\lambda,T)$ and $\lambda_f(T)$ was peformed for Yb:ZBLAN based on the temperature-dependent data obtained from earlier studies [18].

### 3.2 Cooling efficiency modeling

Equipped with the experimental data obtained as described in Section 3.1, $\eta_c(\lambda,T)$ can now be fully calculated. Figure (3) shows the evaluated cooling efficiency using Eq. (1) and the data obtained above. Both Yb:YLF and Yb:ZBLAN are plotted side by side for a direct comparison. Several features of the so-called "cooling window" (blue region) is common to both materials. Cooling window is largest at ambient temperature and contracts (due to diminishing absorption and red-shifting of the mean fluorescence wavelength) upon temperature decrease. A white borderline between cooling (blue) and heating (red) regions corresponds to the spectrum of the minimum achievable temperature, MAT(λ). The drastic difference between two materials is evident at low temperatures. While Yb:ZBLAN cooling window ceases near 190 K, MAT in Yb:YLF is predicted to reach 110 K. Noting the comparable values of external quantum efficiency and background absorption in both materials, this dramatic difference is mainly due to nearly 30 fold enhancement of resonant absorption in YLF, as a consequence of higher doping concentration and, in particular, far less inhomogeneous broadening as compared to amorphous ZBLAN host. Long range order in the crystalline YLF host allows preservation of the oscillator strength of the E4-E5 Stark manifold transition (~ 1020 nm), thus resulting in resonant enhancement of the cooling efficiency. For a given material, MAT depends critically on the value of the background absorption $\alpha_b$. For instance, an order of magnitude improvement in $\alpha_b$ for current Yb:YLF sample would result in MAT near 80 K [19]. We also note that our model predictions of MAT ~ 190 K for excitation near 1015 nm is not consistent with earlier findings [20,14,21], where local cooling down to 77 K with virtually no reduction in cooling efficiency has been reported in Yb:ZBLAN. While it is possible that those samples have been drastically more pure, most of the Yb:ZBLAN samples have been measured to have background absorption $\alpha_b \sim 2 \times 10^{-4}$ cm$^{-1}$ (see for e.g. Ref. [5]).

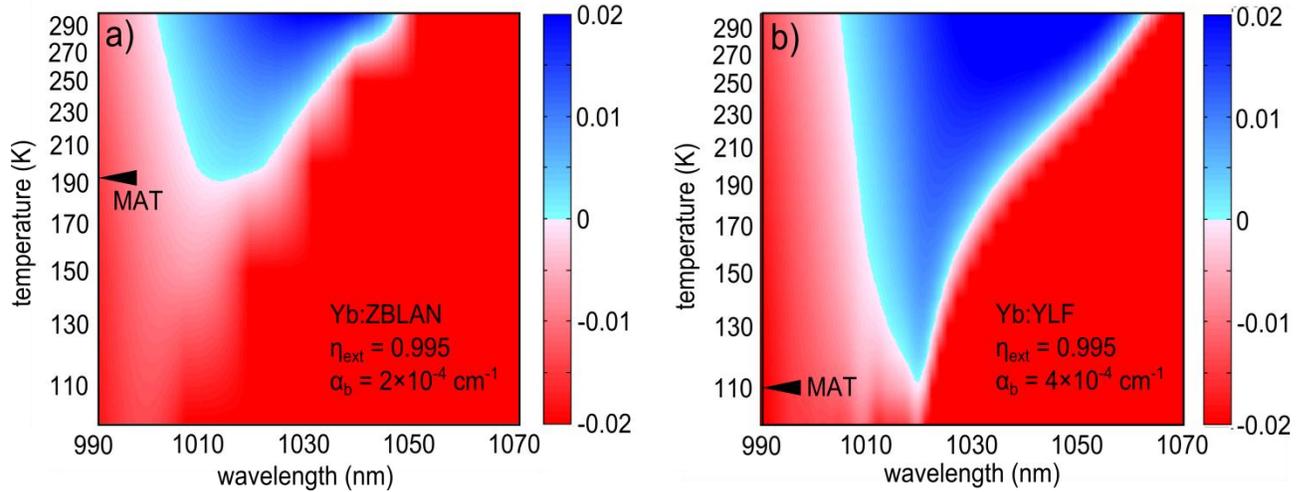

**Figure 3:** Contour plots of cooling efficiency $\eta_c(\lambda, T)$ for Yb:ZBLAN (panel a) and Yb:YLF (panel b) with temperature axis on a logarithmic scale. Transition from cooling (blue) to heating (red) corresponds to MAT(λ) spectrum. Effect of large inhomogeneous broadening in glass host is evident from global MAT in Yb:ZBLAN at ~ 190 K, compared to MAT ~ 110 K for Yb:YLF for otherwise similar parameters of $\eta_{ext}$ and $\alpha_b$.

To verify these predictions and resolve these discrepancies, a highly sensitive and accurate temperature characterization technique is required. In the next sections we present our new technique and discuss measurements on Yb:YLF crystal.

### 3.3 Two-band differential spectral metrology (2B-DSM)

It is important to verify the model predictions (Fig. 3) directly. This requires a sensitive technique that can probe low level of cooling or heating at arbitrary wavelength or temperature. In the following, we detail the implementation of a novel laser calorimetric technique that enabled us to obtain such precision measurements. An initial temperature $T_0$ of the sample is controlled by means of a cryostat, while local temperature deviation $\Delta T$, induced by the tunable pump source, can be ascertained to determine the spectrum of MAT($\lambda$) and hence global minimum achievable temperature. In such arrangement, thermal contact of the cryostat with the sample has to be fine-tuned to be just large enough to maintain sample at $T_0$, while avoiding a large thermal drain that could mask the local temperature deviation $\Delta T$. This is particularly important in crystals with high thermal conductivity (such as YLF), where parasitic heat generated by fluorescence absorption at the thermal contacts inside of a cryostat diffuses back into the laser irradiated region in a short time scale (millisecond or less), and eventually masks the local temperature changes induced by the laser directly. Thus, high sensitivity and fast non-contact temperature probe is desirable.

Techniques such as differential luminescence thermometry [22,23,11] use temperature-dependent spectral change of the analyte to achieve high temperature sensitivity in a non-contact way. However, due to acquisition of the full spectrum, this technique is slow and in addition is not immune to inherent noise from CCD dark current and electronic readout. Here we have developed a technique termed two-band differential spectral metrology (2B-DSM), which utilizes a pair of balanced amplified photodetectors (BAPD) directly in the spectral domain. High common-mode rejection ratio of a BAPD pair allows one to achieve shot-noise limited performance relatively effortlessly.

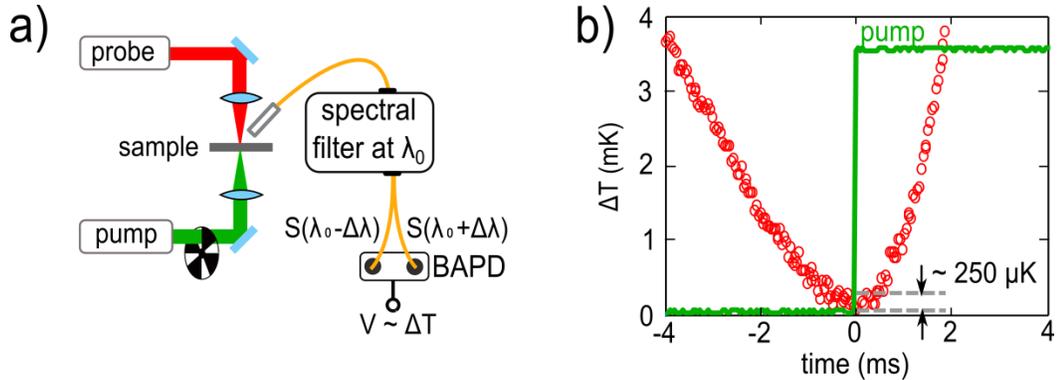

**Figure 4:** (a) Schematic of the pump-probe arrangement for two-band differential spectrum: pump-induced thermal change is interrogated through probe-generated fluorescence $S$, which, after being collected into a fiber is spectrally filtered before a balanced amplified photodetector (BAPD); spectral filter (e.g. a monochromator) centered at $\lambda_0$ followed by BAPD allows for S($\lambda_0 + \Delta\lambda$) − S($\lambda_0 − \Delta\lambda$) ~ ∂S/∂$\lambda$($\lambda_0$) measurement; (b) temperature resolution of 250 µK is demonstrated in GaAs/GaInP double heterostructure.

Pump-probe arrangement for 2B-DSM is shown in Fig. (4), where pump-induced temperature change is interrogated through probe-generated fluorescence spectrum $S(\lambda)$, which is coupled into a fiber and then spectrally filtered before being detected by a pair of balanced amplified photodetectors (BAPD). The role of a spectral filter is to isolate/disperse spectral components of the fluorescence. Such filter for instance can be a monochromator (MC) [24], suitable color filters [25], or any other spectrally selective components. In the case of a MC, portion of the dispersed spectrum at the exit slit is centered at $\lambda_0$, is sampled at two points via optical fibers (Fig. (4)), such that fluorescence spectral slices at $S(\lambda_0 + \Delta\lambda)$ and $S(\lambda_0 − \Delta\lambda)$ are collected (with exit slit width $2\Delta\lambda$), which, after differencing in a BAPD, correspond to signal proportional to ∂S/∂$\lambda$($\lambda_0$). Thermal change of the spectral derivative can thus be measured. An implementation of this technique employed a high quantum efficiency GaAs/GaInP double heterostructure (DHS), heated by 50 mW of 532 nm laser. The luminescence was then induced by a 20 mW laser at 650 nm. While total pump-induced temperature change corresponded to 0.6 degrees, a zoom at the early time reveals 250 μK temperature and sub-millisecond time resolution is possible (Fig. (4b)). We note the generality of this technique and envision applications in differential spectroscopy, flow cytometry, strain-imaging in semiconductors, etc. Next, we apply this technique to laser cooling of Yb:YLF crystal.

### 3.4 Measurement of minimum achievable temperature in Yb:YLF

We applied 2B-DSM to the problem of laser cooling in two separate implementations. In the first implementation, we measured MAT of Yb:YLF, when excited with 1030 nm pump laser (CW Yb:YAG thin disk laser). For this, the 8 W pump was modulated at 10 Hz by means of a mechanical chopper, while 2B-DSM signal was generated with MC grating centered at a wavelength of 996 nm [24].

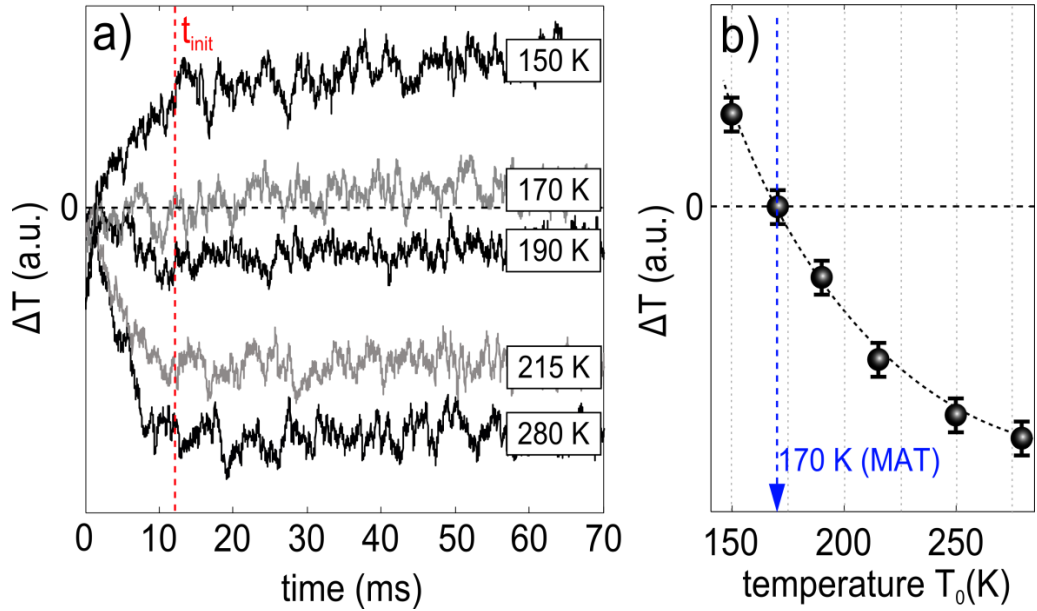

**Figure 5:** (a) time evolution of local temperature deviation $\Delta T$ from global temperature $T_0$ (boxed values) when Yb:YLF is excited at 1030 nm; change from cooling to heating is observed when $T_0$ drops below ∼ 170 K. Temperature dynamics exhibits two time scales, where initially fast temperature change for $t < t_{init}$ signifies local response, while for $t > t_{init}$ contribution of nonlocal heat diffusion is apparent; (b) temperature change $\Delta T$ as measured at $t_{init}$ is plotted versus global temperature $T_0$, with dashed curve serving as an aid to an eye; transition from heating to cooling, i.e. minimum achievable temperature (MAT), is determined to be 170 ± 10 K.

Modulation of 2B-DSM signal during the pump on-time is plotted in Fig. (5a) for several global sample temperatures $T_0$. All time traces show characteristic change in slope around $t_{init}$ ∼ 12-15 ms. For times $t < t_{init}$, 2B-DSM signal exhibits larger slope and change of sign for $T$ above and below 170 K, indicating local temperature evolution, while a slower component and always positive component of the signal at $t > t_{init}$ is indicative of the non-local temperature contribution arising from heat diffusion from the nearby contact points. To avoid such non-local contributions, 2B-DSM signal is plotted at $t_{init}$ in Fig. (5b) versus $T_0$. A clear change from local cooling to heating is observed for $T_0$ = 170 ± 10 K, corresponding to MAT(1030 nm). Resolution of the local temperature difference was determined to be 8 mK, and is mainly limited by relatively temperature insensitive spectrum of the Yb:YLF.

In the second implementation, we bonded the aforementioned GaAs/GaInP DHS directly to the Yb:YLF sample and used a secondary diode laser to probe luminescence of the DHS. This was motivated by high sensitivity of semiconductor bandgap to temperature, with the resolution of 250 µK, as discussed above. Due to large spectral separation of Yb:YLF fluorescence and DHS luminescence, pump-probe can be easily implemented, which in turn allows one to minimize population contributions to the 2B-DSM signal [26]. Higher sensitivity measurement allows us to replace a high power (~ 10 W) Yb:YAG pump with a relatively low power (1.3-1.8 W) broadly tunable CW Ti:Sapphire laser as an excitation source. In this implementation, we were able to measure signal across full MAT spectrum, as presented in Fig. (6).

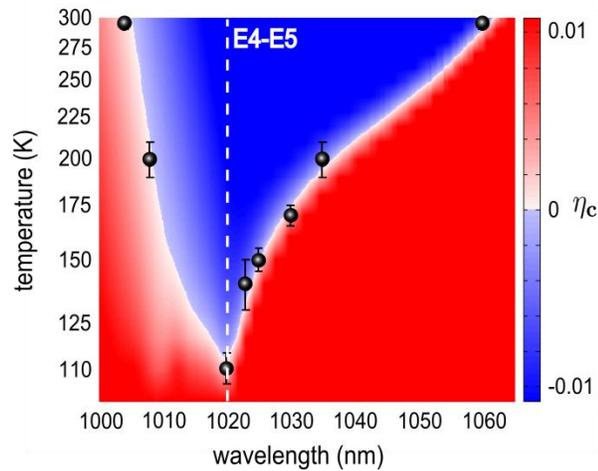

**Figure 6:** Contour plot of the cooling efficiency (Eq. (1)) of Yb:YLF in E∥c orientation. Measured MAT spectrum (black circles) is shown to be in excellent agreement with the model, in particular verifying predicted MAT of 110 K at E4-E5 transition.

Measured MAT spectrum (black circles) is plotted against calculated cooling efficiency and shows excellent quantitative agreement. We note that MAT(1030 nm) = 170 K is reproduced with two independent implementations of the 2B-DSM technique, yielding additional confidence in the data. As a consequence of this agreement, measurement of MAT (1020 nm) = 110 ± 5 K verifies great promise of Yb:YLF current samples to cool at the E4-E5 transition to temperatures below NIST-defined cryogenic onset at 123 K. The demonstrated agreement also verifies the built-in assumptions of the model, namely temperature insensitivity of the background absorption coefficient and the external quantum efficiency. Verified success of the laser cooling model is also important in that calculated MAT of ~ 190 K for Yb:ZBLAN is likely to be an accurate estimate. Several reports exist in the literature on the observation of local cooling below 100 K in ytterbium-doped glasses [20,14,21], all of them using photo-thermal deflection spectroscopy (PTDS) to ascertain local temperature. PTDS relies on a measurement of refractive index change and therefore carrier and thermal contributions to the index change can only be discerned upon optimal spatial separation of pump and probe beams. In general, this separation has to be temperature dependent, for instance to due changes of thermal diffusivity of the material. These considerations render

PTDS results as difficult to interpret unambiguously, especially near the MAT, where local cooling is particularly small. The results presented in this work, however, rely on a spectral differentiation pump-probe scheme (2B-DSM) and hence are only sensitive to thermal contributions. In addition, previous bulk cooling results in Yb:ZBLAN (to 208 K at 1026 nm) [5] and Yb:YLF (to 155 K at 1023 nm) [11] are confirmed in the limit of the respective calculations (Fig. (3)). Demonstrated success of the cooling efficiency model allows us confidently estimate MAT of a typical Yb:ZBLAN to be around 190 K.

## 4. Conclusions

Using temperature-dependent spectroscopic data, we constructed accurate two-dimensional maps of the cooling efficiency $\eta_c(\lambda, T)$ of Yb-doped YLF and ZBLAN using standard cooling model. From this, we predicted minimum achievable temperature (MAT) of current Yb:YLF crystals at 110 K and typical Yb:ZBLAN samples around 190 K. We presented highly-sensitive differential spectrum metrology technique which enables us to verify the cooling efficiency spectra to high accuracy. Most importantly, these measurements have verified the prediction of a minimum achievable temperature of $110 \pm 5$ K at the E4-E5 transition in the Stark manifold near 1020 nm of a 5 mol %-doped Yb:YLF for E||c orientation and $\alpha_b \sim 4x10^{-4}\ cm^{-1}$. This corresponds to the lowest unambiguous measured local temperature in optical refrigeration. With modest improvement in the crystal purity, MAT < 80 K can be expected.

### Acknowledgements

We wish to acknowledge useful discussions with Dr. Markus Hehlen. We also thank Mr. Chengao Wang for GaAs sample preparation and Dr. Michael Hasselbeck for his assistance with LabView software. This work was supported by an AFOSR Multi-University Research Initiative Grant No. FA9550-04-1-0356 entitled Consortium for Laser Cooling in Solids, and a DARPA seedling grant. Research in part was performed while DVS held a National Research Council Research Associateship Award at Air Force Research Laboratory.